\documentclass[12pt]{article}
\usepackage{times}
\usepackage{geometry}
\usepackage{graphicx}
\usepackage{sidecap}
\usepackage[utf8]{inputenc}
\usepackage{enumitem,amssymb}
\usepackage{ragged2e}
\newlist{thematic}{itemize}{8}
\setlist[thematic]{label=$\square$}
\usepackage{pifont}
\newcommand{\cmark}{\ding{51}}%
\newcommand{\done}{\rlap{$\square$}{\raisebox{2pt}{\large\hspace{1pt}\cmark}}%
\hspace{-2.5pt}}

\usepackage[superscript,biblabel]{cite}
\usepackage[super,sort&compress]{natbib}

\usepackage{macros}
\usepackage{aas_macros}

\begin{document}
\thispagestyle{empty}
\raggedright
\huge
Astro2020 Science White Paper \linebreak

Local Dwarf Galaxy Archaeology \linebreak
\normalsize

\noindent \textbf{Thematic Areas:} \hspace*{60pt} $\square$ Planetary Systems \hspace*{10pt} $\square$ Star and Planet Formation \hspace*{20pt}\linebreak
$\square$ Formation and Evolution of Compact Objects \hspace*{31pt} \done Cosmology and Fundamental Physics \linebreak
  $\square$  Stars and Stellar Evolution \hspace*{1pt} \done Resolved Stellar Populations and their Environments \hspace*{40pt} \linebreak
  $\square$    Galaxy Evolution   \hspace*{45pt} $\square$             Multi-Messenger Astronomy and Astrophysics \hspace*{65pt} \linebreak
  
\textbf{Principal Author:}

Name:	Alexander P. Ji
 \linebreak						
Institution:  Carnegie Observatories
 \linebreak
Email: aji@carnegiescience.edu
 \linebreak

\textbf{Co-authors:}\linebreak
Rachael Beaton (Princeton University) \linebreak
Sukanya Chakrabarti (RIT) \linebreak
Gina Duggan (Caltech) \linebreak
Anna Frebel (MIT) \linebreak
Marla Geha (Yale) \linebreak
Matthew Hosek Jr (UCLA)\linebreak
Evan Kirby (Caltech) \linebreak
Ting Li (Fermilab) \linebreak
Ian Roederer (University of Michigan) \linebreak
Joshua Simon (Carnegie Observatories) 
\linebreak 

\justify
\textbf{Abstract:}

Nearby dwarf galaxies are local analogues of high-redshift and metal-poor stellar populations. Most of these systems ceased star formation long ago, but they retain signatures of their past that can be unraveled by detailed study of their resolved stars. Archaeological examination of dwarf galaxies with resolved stellar spectroscopy provides key insights into the first stars and galaxies, galaxy formation in the smallest dark matter halos, stellar populations in the metal-free and metal-poor universe, the nature of the first stellar explosions, and the origin of the elements. Extremely large telescopes with multi-object R=5,000-30,000 spectroscopy are needed to enable such studies for galaxies of different luminosities throughout the Local Group.

\pagebreak
\clearpage
\setcounter{page}{1}
\section{A Window to the First Stars and Galaxies}

Observationally accessing the era of first stars and galaxies ($40 > z > 6$) is a fundamental goal of modern astrophysics.
Local galaxies contain intact stellar populations that provide an archaeological record of star formation and chemical enrichment in faint galaxies at high redshift, a regime that is difficult or inaccessible to direct observation \cite{BoylanKolchin16}.  Nearby {\it dwarf} galaxies are particularly attractive for studying near-field cosmology, star formation histories, and chemical evolution.
The surviving stars in these dwarf galaxies are local analogues of high-redshift and metal-poor stellar populations, while their chemical abundances retain signatures of massive stars from that early epoch.
Dwarf galaxies' small masses and typically simple star formation histories also enhance the signatures of many important physical processes, such as stellar feedback\cite{Wheeler14}, gas inflows and outflows\cite{Kirby11}, chemical yields from rare events\cite{Ji16}, and enrichment from metal-free stars\cite{Jeon17}.

The past decade has seen an explosion in the number of detected dwarf galaxies, including the lowest luminosity and surface brightness galaxies known in the universe \cite{Bechtol15,Koposov15,vanDokkum15,Torrealba18}, and a whole population of galaxies around M31 \cite{Martin13}.
LSST is expected to more than double the number of known dwarf galaxies \cite{simon_dwgal_wp,lsst_wp}.
This expanding sample of dwarf galaxies provides the opportunity to study universal physical processes in galaxy formation and chemical evolution, as well as deviations from these trends (e.g., the impact of a major merger or rare stellar explosion).
Already with the current sample of galaxies, we have 
discovered new and different abundance patterns and trends \cite{Venn12,McWilliam13,Frebel14,Ji16,Roederer16,McWilliam18};
pushed beyond the Milky Way's virial radius \cite{Leaman12,Tollerud12,Vargas14,Escala18,Kirby17,Taibi18}
and into the faintest galaxies known \cite{Simon11,Vargas13,Martin16,Kirby17b};
studied chemodynamic sub-populations and radial gradients\cite{Koposov11,Lemasle12,Lemasle14,Kordopatis16};
and combined information in color-magnitude diagrams and spectra to study star formation histories and age-metallicity relations \cite{Weisz14,Savino15,Brown14,Norris17,Hill18}.
These efforts have revealed that local dwarf galaxies display a diversity of star formation and chemical enrichment histories.
However, detailed spectroscopic study of resolved stellar populations in dwarf galaxies is still mostly limited to a handful of relatively luminous or nearby Milky Way satellites.
If we could spectroscopically survey the entire Local Group with similar fidelity, that would provide a representative picture of stars and galaxies at $z > 6$ \cite{BoylanKolchin16}, while also sampling a wide range of galaxy luminosities and environments.
The construction of Extremely Large Telescopes (ELTs) in the next decade will enable such archaeological reconstruction of the formation histories and stellar populations of dwarf galaxies across the whole Local Group.

\section{Key Questions}
We focus on key questions answerable in the coming decade that require spectroscopy and elemental abundances, augmenting a white paper on near-field cosmology \cite{weiszmbk_wp} and complementing topics covered by other white papers on new dwarf galaxy discoveries \cite{lsst_wp,simon_dwgal_wp}, dwarf galaxy kinematics \cite{simon_dwgal_wp,li_mse}, stellar halos and streams \cite{li_mse,sanderson_wp}, and Pop\,III star signatures \cite{roederer_ksp}.

\begin{SCfigure}
    \includegraphics[width=10cm]{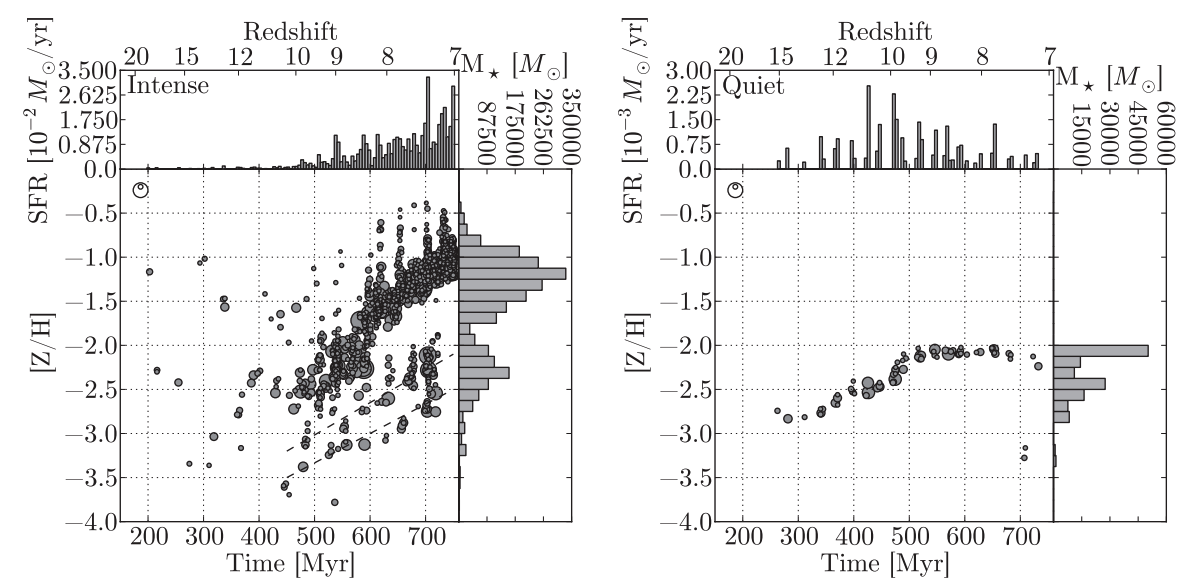}
    \caption{Simulated age-metallicity relations of two galaxies forming before reionization \cite{Wise12}.
    Galaxies in this epoch can contain lots of interesting structure, but the stellar ages are entirely unresolved by current observations.
    Detailed chemical abundances of stars forming in and after the epoch of reionization provide information that can help understand the nature of early galaxy formation.
    }
    \label{fig:sfr}
\end{SCfigure}

\vspace{3mm}
\textbf{What is the threshold of galaxy formation before and after reionization?}
Theoretical arguments suggest star formation can occur in dark matter halos of mass above $10^{5.5-8} M_\odot$ prior to reionization and $\gtrsim 10^{9} M_\odot$ after reionization (Fig~\ref{fig:sfr}) \cite{Bullock00,Gnedin00,Bromm11,Wise12,Schauer19}.
These mass thresholds directly determine which galaxies contribute to reionization, and the number abundance of dwarf galaxies in the Local Group \cite{weiszmbk_wp}.
Current observations suggest that galaxies with stellar mass $\lesssim 10^6 M_\odot$ are near the threshold where reionization is important \cite{Brown14,Weisz14b}, but further progress requires understanding the star formation and gas accretion histories of faint galaxies in the first Gyr of the universe.

\begin{figure}
    \centering
    \includegraphics[width=\linewidth]{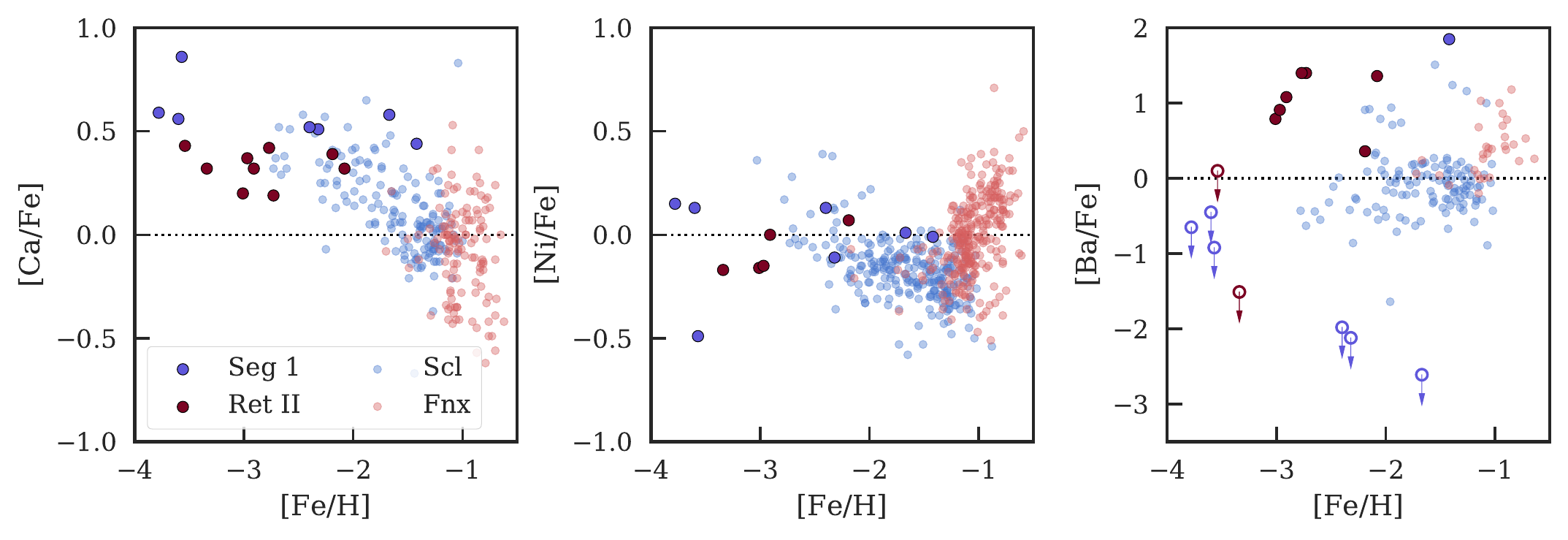}
    \caption{Chemical evolution trends for two ultra-faint dwarf galaxies from high-resolution spectroscopy (Segue 1, Reticulum II)\cite{Frebel14,Ji16} and two classical dSphs from medium-resolution spectroscopy (Sculptor, Fornax) \cite{Kirby11,Kirby18,Duggan18}.
    $\alpha$-elements like Ca constrain the star formation history of the galaxy.
    Fe-peak elements like Ni are related to the Type Ia explosion mechanism.
    Neutron-capture elements like Ba can be used to understand the origin of the heaviest elements.
    Being luminous and nearby, Scl and Fnx are two of the best-studied dwarf galaxies, with well-populated abundance trends.
    ELTs are needed for similar studies in the faintest galaxies and outside the Milky Way.
    }
    \label{fig:chemevol}
\end{figure}

Accessing this earliest era benefits greatly from combining color-magnitude diagrams (CMDs) and spectroscopic chemical abundance information.
Spectroscopic abundances help break the age--metallicity degeneracy in CMDs\cite{Brown14}, while the star formation histories provided by CMDs aid the interpretation of chemical evolution models\cite{Gilmore91,Kirby17,Hill18}.
$\alpha$-element abundances indicate at what metallicity Type Ia supernovae begin to contribute metals, which has been extensively used to estimate typical galaxy formation timescales (Fig~\ref{fig:chemevol})\cite{Gilmore91,Kirby11,Frebel14}.
Other delayed enrichment sources, such as asymptotic giant branch (AGB) stars or neutron star mergers, can also be used to determine characteristic timescales for star formation in dwarf galaxies \cite{Duggan18, Hill18}.
Chemical abundances also constrain the amount of gas accretion in a galaxy \cite{Kirby13}.
Joint consideration of CMDs and abundances for star formation histories is still relatively new, and so far it has only been applied to very luminous or very nearby Milky Way satellites \cite{Lemasle12,Lemasle14,Brown14,Norris17,Hill18,Taibi18}.
Large telescopes will enable such detailed studies in the faintest known galaxies, where the signatures of reionization may be most prominent, as well as in field galaxies and M31 satellites.

\vspace{3mm}
\textbf{Are distinct chemodynamical populations present in all dwarf galaxies?}
Dwarf galaxies are simple when compared to the Milky Way, but they still contain features including complex stellar populations, radial gradients, faint stellar halos, and other signatures of hierarchical galaxy formation (e.g. Fig~\ref{fig:sfr}).
Such signatures have only been observed in a few of the most luminous nearby galaxies \cite{Koposov11, Weisz14, Kordopatis16}, and the vast majority are still modeled as single homogeneously evolving systems.
A key limitation thus far is that spectroscopy of hundreds of stars per galaxy are needed to robustly detect and characterize multiple chemodynamic populations, requiring larger spatial coverage and deeper observations with multi-object spectrographs.
For the closest dwarf galaxies ($\lesssim 50$ kpc), spectroscopy with future large telescopes can feasibly reach the main sequence turnoff, providing an opportunity to obtain both precise ages and metallicities for individual stars.
This approach offers the potential to resolve the earliest signs of hierarchical galaxy formation.

\vspace{3mm}
\textbf{What are the sources and timescales of production of different elements?}
The lower star formation efficiencies in dwarf galaxies amplify the impact of time-delayed nucleosynthetic sources on stellar abundances (Fig~\ref{fig:chemevol}). We highlight the explosion mechanism of Type Ia supernovae and the origin of neutron capture elements as two of the most interesting open questions.

\emph{What is the explosion mechanism for Type~Ia supernovae?}
Despite the importance of Type~Ia supernovae for cosmology\cite{Riess98} and chemical evolution\cite{Gilmore91}, there is not yet a consensus on the physical mechanism of the explosion.  The progenitor is clearly a binary system that includes at least one white dwarf.  However, the system could contain one or two white dwarfs (single- or double-degenerate).  The common wisdom that Type~Ia supernovae explode by exceeding the Chandrasekhar mass has been challenged by theory\cite{Shen17} and observations\cite{Kerzendorf09}.  It has been suggested that Fe-peak elements, like Mn and Ni, can at least distinguish the mass of the exploding white dwarf\citep{Seitenzahl13, McWilliam18}.  In the next decade, further improvements in theoretical modeling and precision abundance measurements may even lead to a definitive answer on the question of single- or double-degenerate progenitors.

\emph{How are neutron-capture elements created?} 
Neutron-capture elements like Sr, Ba, and Eu are synthesized primarily in the $s$- and $r$-processes.
AGB stars have been identified as the dominant site of the $s$-process, but our understanding of their chemical contribution at zero or low metallicity suffers from both limited observational data and theoretical predictions\cite{Karakas16,Frischknecht16}.
The site(s) of the $r$-process is likely compact object mergers \cite{Lattimer77,LIGOGW170817b} and/or the death of massive stars\cite{Mosta17,Siegel18}, but the relative dominance of these two channels is still debated \cite{Cote18,Siegel18,Duggan18,Hill18}.
Dwarf galaxies have diverse but simple star formation histories, allowing them to provide insight into the yields, delay time distributions, metallicity dependence, and relative occurrence rates of different neutron-capture element sources.

\begin{figure}
    \centering
    \includegraphics[width=14cm]{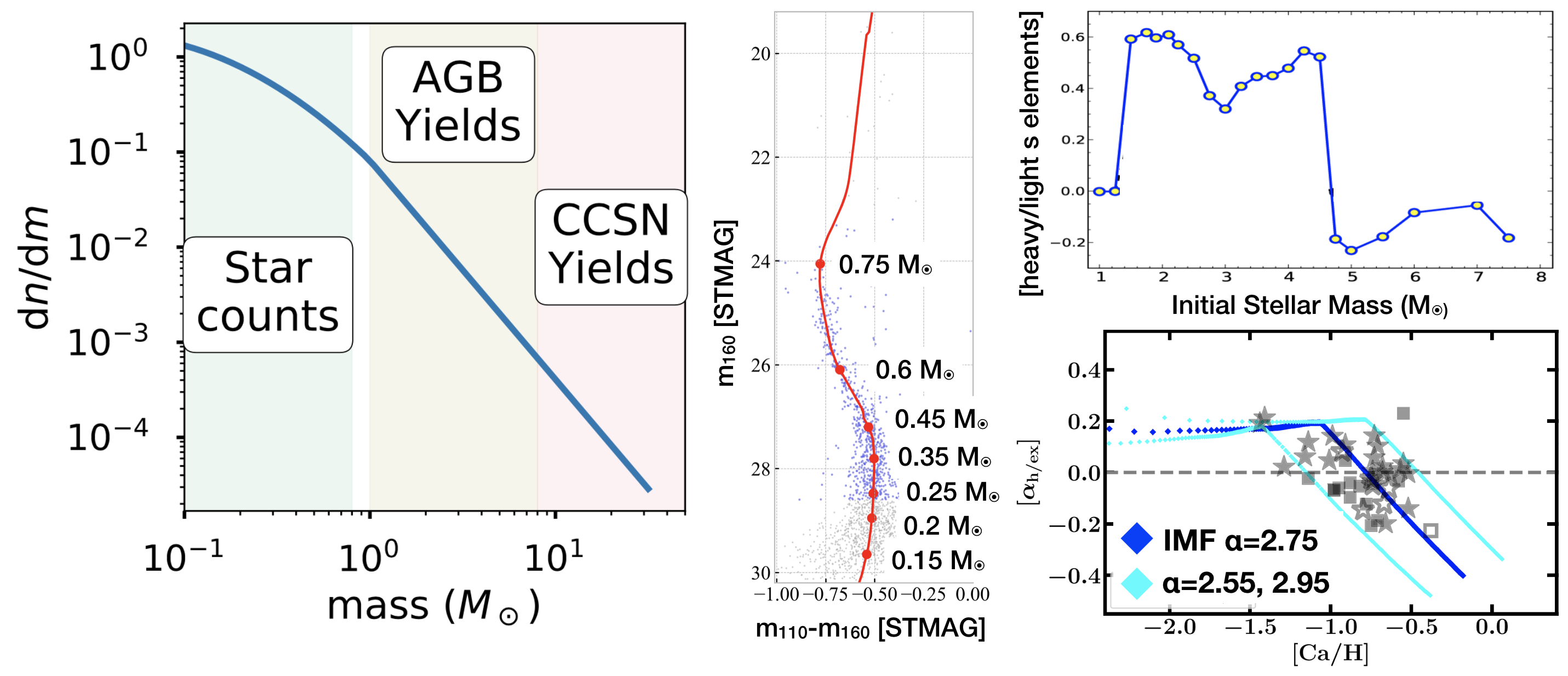}
    \caption{Constraining the IMF with resolved stellar populations. Left: Chabrier IMF vs mass. Center: stars with mass $< 0.8 M_\odot$ can be directly counted \cite{Gennaro18}. Top right: $1-8 M_\odot$ stars release different $s$-process yields as AGB stars \cite{Karakas16}.
    Bottom right: massive stars produce different ratios of $\alpha$-elements, resulting in different abundance trends depending on IMF slope \cite{Carlin18}.
    }
    \label{fig:imf}
\end{figure}

\vspace{3mm}
\textbf{How do the IMF and binary fraction vary across environment?}
The initial mass function (IMF) is a fundamental parameter of star and galaxy formation that is often assumed to be universal and unchanging.
However, in the past decade there has been mounting evidence that the IMF changes in different galactic environments \cite{Conroy12, Geha13, McWilliam13}.
At the low mass end ($<0.8 M_\odot$), direct star counts in dwarf galaxies show that the IMF appears to be bottom light, although with variations between galaxies \cite{Geha13, Gennaro18}.
At the high mass end ($>8 M_\odot$), abundance ratios reflecting the initial masses of supernovae (such as [Mg/Ca]) also appear to vary amongst dwarf galaxies \cite{McWilliam13,Carlin18,Hill18}.
Variations in the intermediate mass range ($1-8 M_\odot$) have not been studied yet, but can potentially be probed by mass-dependent $s$-process nucleosynthesis in AGB stars \cite{Karakas16}.
In the next decade, we have the opportunity to probe the full extent of IMF variation in different galactic environments by combining these probes (Fig~\ref{fig:imf}).
The low mass end is limited by stellar crowding and sample contamination by unresolved background galaxies. High spatial resolution photometry with JWST is needed to robustly constrain the shape of the IMF down to its expected peak at $\sim$0.2 M$_{\odot}$  \cite{Gennaro18,ElBadry17}. The high mass end is limited by abundance precision and sample sizes, requiring large samples of stars with high-resolution spectroscopy \cite{McWilliam13,Carlin18,Hill18}.

In dwarf galaxies, the prevalence of binary stars and their orbital properties could be significantly different from Milky Way stellar populations.
Several studies suggest that metal-poor stars in the Milky Way and dwarf galaxies appear to have higher binary fractions and different period distributions compared to more metal-rich field stars \cite{Martinez11,Minor18,Moe18,Spencer18}.
Confirming this result could have important implications for IMFs derived from star counts \cite{Gennaro18}, ionizing photon production \cite{Ma16}, high-redshift galaxy emission line ratios \cite{Strom17}, and binary star nucleosynthesis (e.g., Type Ia supernovae, neutron star mergers).
Multi-epoch radial velocity observations and large searches for carbon-enhanced stars (which reveal unseen binary companions through binary mass transfer) in these dwarf galaxies are needed to understand whether binary fractions change across galactic environment.

\vspace{3mm}
\textbf{What signatures of metal-free stars are present in dwarf galaxies?}
The first metal-free (Pop~III) stars are unusually massive and likely produced unique abundance signatures in their supernova explosions \cite{Bromm13}.
Stellar abundances of extremely metal-poor stars in the ultra-faint dwarf galaxies ($M_V > -7.7$) are predicted to be relatively pure tracers of these explosions \cite{Ji15, Salvadori15, Jeon17}. 
Fitting the abundances of these stars to supernova model yields can provide insight into the mass distribution of the first stars \cite{Ishigaki18}, but the total number of such stars in dwarf galaxies is still relatively small.
It is also hypothesized that some metal-free stars have low enough masses to survive until the present day \cite{Stacy14}, which could comprise as much as 1\% of stars in faint dwarf galaxies \cite{Magg18}.
Understanding and searching for signatures of metal-free stars requires determining abundances of $\sim 10^{2}$ stars in a large number of faint dwarf galaxies.

\section{Summary and Requirements for the Next Decades}
Spectroscopy of individual stars in dwarf galaxies, obtained with 6-10~m telescopes over the last few decades, has provided critical insights into a wide range of scientific topics.
Detailed study of these stellar populations may be the only way to access the internal properties of typical star-forming galaxies at $z > 6$.
However, individual stars at extragalactic distances are very faint, requiring multi-hour integrations on the largest telescopes to measure the weak absorption features in these metal-poor stars.
With medium resolution spectroscopy ($R \gtrsim 5000$), spectral synthesis techniques can provide the majority of elements of interest with a precision ${\sim}0.2$\,dex \cite{Kirby11,Kirby18,Duggan18,Escala18}.
High-resolution spectroscopy ($R \gtrsim 20,000$), however, is still needed to detect several key elements (like Eu) and most metals in the lowest-metallicity stars ($\mbox{[Fe/H]} \lesssim -3$), where metal lines become too weak to detect with low and moderate resolution spectroscopy \cite{roederer_ksp}.

\begin{SCfigure}
    \centering
    \includegraphics[width=8cm]{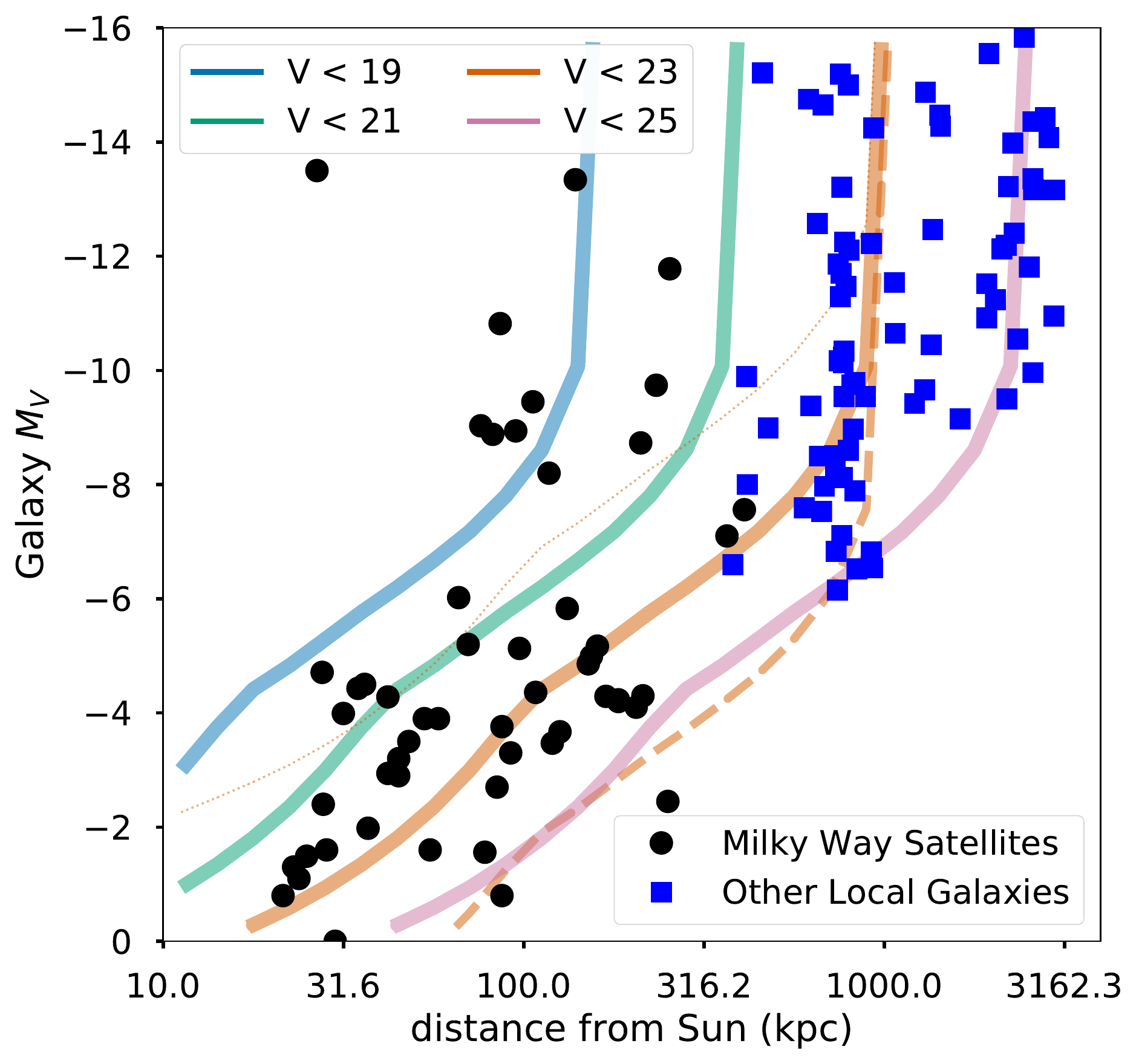}
    \caption{Luminosity vs distance for Local Group galaxies \cite{Simon19,McConnachie12}.
    Galaxies to the upper-left of the colored lines have ${>}100$ stars available for spectroscopy at a given magnitude limit (estimated with a 12 Gyr metal-poor isochrone\cite{Dotter08}).
    The dashed and dotted lines for $V < 23$ indicate the limits for 10 and 1000 stars, respectively.
    For high-resolution spectroscopy, Magellan/MIKE can measure abundances of 10--20 elements for stars at $V\sim19$ in 6-8h of integration \cite{Frebel14}.
    For medium-resolution spectroscopy, Keck/DEIMOS can measure [Fe/H] and [$\alpha$/Fe] for stars at $V \sim 22$ in 6h of integration \cite{Vargas14,Escala18}.
    ELTs will increase observation depth by 2 magnitudes, allowing high-resolution abundances for 10--100 stars out to the virial radius of the Milky Way,
    and medium-resolution abundances of ${>}100$ stars out M31 and beyond.
    }
    \label{fig:surveylimit}
\end{SCfigure}

Fig~\ref{fig:surveylimit} shows all \emph{currently known} dwarf galaxies in the Local Group.
The colored lines indicate which galaxies have ${>}100$ stars available for spectroscopy at a given magnitude limit.
The number of faint galaxies ($M_V > -8$) is expected to grow substantially in the coming decade \cite{simon_dwgal_wp}.
ELTs would provide 2 extra magnitudes of depth, paying huge dividends for dwarf galaxies by enabling spectroscopic abundances of statistically meaningful samples of stars out beyond M31 and in all nearby ultra-faint dwarf galaxies.
Achieving such sample sizes will require multi-object spectroscopy at both medium- and high-resolution, though we note that the stellar densities of most galaxies are low enough that adaptive optics is not required.
The moderate field-of-views planned for multi-object spectrographs on ELTs (25--50~arcmin$^{2}$) are also sufficient to cover known galaxies, although closer Milky Way satellites benefit from wider fields.
Thus, \textbf{we recommend the construction of ELTs with multi-object spectrographs to enable archaeological study of dwarf galaxy formation histories across the Local Group.}

\newpage

\bibliographystyle{JHEP}

\providecommand{\href}[2]{#2}\begingroup\raggedright\endgroup

\end{document}